\documentclass[preprint,aps,pra,showpacs,onecolumn]{revtex4}
\usepackage{graphicx}
\usepackage{mathrsfs}
\usepackage{bbm}
\usepackage{amssymb}
\usepackage{amsmath}
\usepackage{subfigure}
\usepackage{slashbox}
\usepackage{bm}
\usepackage{txfonts}
\newcommand{\ket}[1]{|#1\rangle}
\newcommand{\bra}[1]{\langle#1|}

\newcommand{\lr}\longrightarrow
\newcommand{\ra}\rightarrow

\begin{document}

\title{Yang-Baxter $\breve{R}$ matrix, Entanglement and Yangian}

\author{Gangcheng Wang}
\author{Kang Xue}%
\email{youngcicada@163.com}
\author{Chunfang Sun}
\author{Guijiao Du}
\affiliation{%
School of Physics, Northeast Normal University\\ Changchun 130024,
People's Republic of China }

\date{\today}

\begin{abstract}
We present a method to construct ``X" form unitary Yang-Baxter
$\breve{R}$ matrices, which act on the tensor product space
$V_{i}^{j_{1}}\otimes V_{i+1}^{j_{2}}$. We can obtain a set of
entangled states for $(2j_{1}+1)\times (2j_{2}+1)$-dimensional
system with these Yang-Baxter $\breve{R}$ matrices. By means of
Yang-Baxter approach, a $8\times 8$ Yang-Baxter Hamiltonian is
constructed. Yangian symmetry and Yangian generators as shift
operators for this Yang-Baxter system are investigated in detail.
\end{abstract}


\maketitle

\section{\label{intro}Introduction}
Quantum entanglement\cite{ben1,ben2,ben3,murao}, which is a bizarre
of quantum theory, has been recognized as an important resource for
applications in quantum information and quantum computation
processing. Quantum gates\cite{nielsen} are represented by unitary
matrices, and they are building blocks of a quantum computer. On the
other hand, the topological quantum computation(TQC) also has been
studied by researchers\cite{nayak}. Thus quantum computation is one
of the important approaches to achieve a fault-tolerant quantum
computer. This proposal relies on the existence of topological
states of matter, whose quasiparticle excitations are non-Abelian
anyons. Thus quasiparticles obey non-Abelian braiding statistics,
and quantum gate operators are implemented by braiding
quasiparticles.

Recently, Kauffman \emph{et.al.} have shown that topological
entanglement and quantum entanglement have deep
relations\cite{kauffman1,kauffman2,zhang1}. The authors propose that
it is more fundamental to view braid matrices(or solutions to
Quantum Yang-Baxter Equation\cite{yang,baxter}), which can implement
topological entanglement, as universal quantum gates. For example,
the authors showed that the Bell matrix is nothing, but a braid
matrix, and thus braid matrix local equivalent to a
Control-Not(CNOT) gate\cite{kauffman2}. This motivated a novel way
to study quantum entanglement by means of Yang-Baxter
approach\cite{zhang2,chen1,chen2,chen3,hu1,hu2,wang1}.

The Yangian theory established by Drinfeld offer a mathematic method
for the studies about the symmetry of quantum integrable models in
physics\cite{drin}. Many researchers have explored the role of
Yangian operators in physics\cite{xue1,xue2,xue3}. For example, by
means of Yangian, we can investigate the symmetry for the integrable
systems and shift operators. But many researchers worked on complex
systems, this motivated us to search a simple system with Yangian
symmetry to investigate the role of Yangian operators in this
system.

In Sec.~\ref{sec2}, we will present a method for constructing the
``X" form Yang-Baxter $\breve{R}$ matrices, and then we will
investigate the entanglement properties in Sec.~\ref{sec3}. In
Sec.\ref{sec4}, we construct Yang-Baxter Hamiltonian with a
$8\times8$ $``X"$ form Yang-Baxter $\breve{R}$ matrix, then Yangian
symmetry and shift operators are studied in this Yang-Baxter system.

\section{\label{sec2}The ``X" form Yang-Baxter $\breve{R}$ matrices}
In this paper, Yang-Baxter $\breve{R}^{j_{1}j_{2}}(\theta)$ matrix
and $M^{j_{1}j_{2}}$ matrix are $(2j_{1}+1)\times
(2j_{2}+1)$-dimensional matrices acting on the tensor product
$V^{j_{1}}\otimes V^{j_{2}}$, where $V^{j_{1}}$ and $V^{j_{2}}$ are
$(2j_{1}+1)$ and $(2j_{2}+1)$ dimensional vector space,
respectively. As Yang-Baxter $\breve{R}^{j_{1}j_{2}}(\theta)$ matrix
and $M^{j_{1}j_{2}}$ matrix acting on the tensor product
$V_{i}^{j_{1}}\otimes V_{i+1}^{j_{2}}$, we denote them by
$\breve{R}_{i}^{j_{1}j_{2}}(\theta)$ and $M_{i}^{j_{1}j_{2}}$,
respectively. The notation $I^{j_{1}j_{2}}$ denotes
$(2j_{1}+1)\times (2j_{2}+1)$-dimension identity matrix.

 Let matrices $M^{j_{1}j_{2}}$ and $M^{j_{2}j_{1}}$ satisfying the
following relations,
\begin{eqnarray}\label{m-relation}
\begin{array}{ll}
  [M^{j_{1}j_{2}}]^{2} =[M^{j_{2}j_{1}}]^{2} = I^{j_{1}j_{2}} &  \\
  M_{12}^{j_{1}j_{2}}M_{23}^{j_{2}j_{1}} = M_{23}^{j_{2}j_{1}}M_{12}^{j_{1}j_{2}}, & (i.e.
 [M_{12}^{j_{1}j_{2}},M_{23}^{j_{2}j_{1}}]=0) \\
  M_{12}^{j_{2}j_{1}}M_{23}^{j_{1}j_{2}} = M_{23}^{j_{1}j_{2}}M_{12}^{j_{2}j_{1}}, & (i.e.
 [M_{12}^{j_{2}j_{1}},M_{23}^{j_{1}j_{2}}]=0).
\end{array}
\end{eqnarray}
In this paper, we set
$[M^{j_{1}j_{2}}]^{a\alpha}_{b\beta}=[M^{j_{2}j_{1}}]^{\alpha
a}_{\beta b}$($-j_{1}\leq a,b \leq j_{1}$ and $-j_{2}\leq
\alpha,\beta\leq j_{2}$)for convenience.
 Then two spectral-dependent Yang-Baxter $\breve{R}$ matrices via Yang-Baxterization\cite{xue4,xue5,xue6} is obtained to be,
\begin{eqnarray}\label{r-matrix}
\begin{array}{l}
  \breve{R}^{j_{1}j_{2}}(\theta)=e^{-i\frac{\theta}{2}M^{j_{1}j_{2}}}=cos\frac{\theta}{2}I^{j_{1}j_{2}}-isin\frac{\theta}{2}M^{j_{1}j_{2}}, \\
  \breve{R}^{j_{2}j_{1}}(\theta)=e^{-i\frac{\theta}{2}M^{j_{2}j_{1}}}=cos\frac{\theta}{2}I^{j_{1}j_{2}}-isin\frac{\theta}{2}M^{j_{2}j_{1}}.
\end{array}
\end{eqnarray}
Here we used Tayloy expansion to derive the right hand of
Eq.~\ref{r-matrix}. If the matrices $M^{j_{1}j_{2}}$ and
$M^{j_{2}j_{1}}$ are Hermitian matrices(\emph{i.e.}
$[M^{j_{1}j_{2}}]^{\dag} =M^{j_{1}j_{2}}$ and
$[M^{j_{2}j_{1}}]^{\dag} = M^{j_{2}j_{1}}$), then we can verify that
the matrices $\breve{R}^{j_{1}j_{2}}$ and $\breve{R}^{j_{2}j_{1}}$
are unitary(\emph{i.e.}
$\breve{R}^{j_{1}j_{2}}(\theta)^{\dag}\breve{R}^{j_{1}j_{2}}(\theta)=\breve{R}^{j_{1}j_{2}}(\theta)\breve{R}^{j_{1}j_{2}}(\theta)^{\dag}=I^{j_{1}j_{2}}$
and
$\breve{R}^{j_{2}j_{1}}(\theta)^{\dag}\breve{R}^{j_{2}j_{1}}(\theta)=\breve{R}^{j_{2}j_{1}}(\theta)\breve{R}^{j_{2}j_{1}}(\theta)^{\dag}=I^{j_{2}j_{1}}$).

We can easily prove that $\breve{R}^{j_{1}j_{2}}(\theta)$ and
$\breve{R}^{j_{2}j_{1}}(\theta)$ satisfy the following Yang-Baxter
equation(YBE),
\begin{eqnarray}\label{ybe}
\begin{array}{l}
  \breve{R}_{12}^{j_{1}j_{2}}(\theta_{1})\breve{R}_{23}^{j_{2}j_{1}}(\theta_{1}+\theta_{2})\breve{R}_{12}^{j_{1}j_{2}}(\theta_{2})=\breve{R}_{23}^{j_{2}j_{1}}(\theta_{2})\breve{R}_{12}^{j_{1}j_{2}}(\theta_{1}+\theta_{2})\breve{R}_{23}^{j_{2}j_{1}}(\theta_{1}), \\
  \breve{R}_{12}^{j_{2}j_{1}}(\theta_{1})\breve{R}_{23}^{j_{1}j_{2}}(\theta_{1}+\theta_{2})\breve{R}_{12}^{j_{2}j_{1}}(\theta_{2})=\breve{R}_{23}^{j_{1}j_{2}}(\theta_{2})\breve{R}_{12}^{j_{2}j_{1}}(\theta_{1}+\theta_{2})\breve{R}_{23}^{j_{1}j_{2}}(\theta_{1}).
\end{array}
\end{eqnarray}
where parameters $\theta_{1}$ and $\theta_{2}$ are called as
spectral parameters. For convenience, we take $M^{j_{1}j_{2}}$ and
$M^{j_{2}j_{1}}$ as $[M^{j_{2}j_{1}}]^{\alpha a}_{\beta
b}=[M^{j_{1}j_{2}}]^{a\alpha}_{b\beta}=e^{-i\varphi_{a\alpha}}\delta_{a,-b}\delta_{\alpha,-\beta}$.
Considering the first equation in Eqs.~\ref{m-relation}, we set
$\varphi_{a\alpha}=-\varphi_{-a-\alpha}$. Substituting
$M^{j_{1}j_{2}}$ and $M^{j_{2}j_{1}}$ into the second and the third
relations in Eqs.~\ref{m-relation}, we can obtain the following
conditions,
\begin{eqnarray}\label{cond1}
\begin{array}{c}
  \varphi_{a\alpha}+\varphi_{-a\alpha}=\varphi_{b\alpha}+ \varphi_{-b\alpha},\\
    \varphi_{a\alpha}+\varphi_{a-\alpha}=\varphi_{a\beta}+ \varphi_{a-\beta}.
\end{array}
\end{eqnarray}
With this method, we can obtain high dimentional Yang-Baxter
$\breve{R}^{j_{1}j_{2}}$ matrices easily. By means of these
Yang-Baxter $\breve{R}^{j_{1}j_{2}}$ matrices, we can investigate
quantum entanglement consequently.
\section{\label{sec3}The ``X" form $\breve{R}$ matrices as quantum gates}
In this section, three examples are shown to illustrate this method
in detail. The case $j_{1}=j_{2}=1/2$ gives us a $4\times 4$ unitary
Yang-Baxter $\breve{R}^{1/2,1/2}(\theta)$ matrix. Thus we can view
the $\breve{R}^{1/2,1/2}(\theta)$ matrix as a quantum gate for
two-qubit system. If $j_{1}=1$ and $j_{2}=1/2$, we can obtain a $6
\times 6$ Yang-Baxter $\breve{R}^{1,1/2}$ matrix. This unitary
$\breve{R}^{1,1/2}$ can entangle quantum states in system with one
qubit and one qutrit. When $j_{1}=3/2$ and $j_{2}=1/2$, a
three-qubit quantum gate $\breve{R}^{3/2,1/2}$ can be obtained. For
quantify the entanglement of bi-particle system states, we use the
negativity\cite{zycz,wangxg} defined by,
\begin{eqnarray}\label{neg}
N(\rho) &=& \frac{\|\rho^{T_{B}}\|_{1}-1}{d-1}.
\end{eqnarray}
where $\rho^{T_{B}}$ is the partial transpose of a state $\rho$ in
$d\times d'$$(d\leq d')$ quantum system, and the notation
$\|A\|_{1}=Tr\sqrt{A^{\dagger}A}$ denotes the trace norm of $A$. It
should be noted that the negativity criterion is necessary and
sufficient only for $2 \otimes 2$ and $2 \otimes 3$ quantum systems.
\subsection{The $4 \times 4$ ``X" form $\breve{R}$ matrix}
If $j_{1}=j_{2}=1/2$, the equations in Eqs.~(\ref{m-relation}) can
be simplified as $[M^{1/2,1/2}]^{2}=I^{1/2,1/2}$ and
$[M^{1/2,1/2}_{12},M^{1/2,1/2}_{23}]=0$. Then we can obtain a matrix
$M^{1/2,1/2}$ as following,
\begin{equation*}
    M^{1/2,1/2}=e^{-i(\varphi+\frac{\pi}{2})}s_{1}^{+}s_{2}^{+}+s_{1}^{+}s_{2}^{-}+s_{1}^{-}s_{2}^{+}+e^{i(\varphi+\frac{\pi}{2})}s_{1}^{-}s_{2}^{-}.
\end{equation*}
The Yang-Baxter $\breve{R}^{1/2,1/2}$ matrix can be obtained as
follows,
\begin{eqnarray*}
  \breve{R}^{1/2,1/2}(\theta)=e^{-i\frac{\theta}{2}M^{1/2,1/2}}=\cos\frac{\theta}{2}I^{1/2,1/2}-i\sin\frac{\theta}{2}M^{1/2,1/2},
\end{eqnarray*}
or in matrix form,
\begin{eqnarray*}
  \breve{R}^{1/2,1/2}(\theta) &=& \left(
                    \begin{array}{cccc}
                      \cos\frac{\theta}{2} & 0 & 0 & -\sin\frac{\theta}{2}e^{-i\varphi} \\
                      0 & \cos\frac{\theta}{2} & -i\sin\frac{\theta}{2} & 0 \\
                      0 & -i\sin\frac{\theta}{2} & \cos\frac{\theta}{2} & 0 \\
                      \sin\frac{\theta}{2}e^{i\varphi} & 0 & 0 & \cos\frac{\theta}{2} \\
                    \end{array}
                  \right).
\end{eqnarray*}
In this section, we choose $\{\ket{00}, \ket{01}, \ket{10},
\ket{11}\}$ as standard bases. Acting this Yang-Baxter
$\breve{R}^{1/2,1/2}$ matrix on the standard bases, we can obtain a
set entangled states $\{\ket{e_{i}},i=1,2,3,4\}$ㄛ
\begin{eqnarray*}
  \left(
     \begin{array}{c}
       \ket{e_{1}} \\
       \ket{e_{2}} \\
       \ket{e_{3}} \\
       \ket{e_{4}} \\
     \end{array}
   \right)
   = \breve{R}^{1/21/2}(\theta)\left(
                                 \begin{array}{c}
                                   \ket{00} \\
                                   \ket{01} \\
                                   \ket{10}\\
                                   \ket{11} \\
                                 \end{array}
                               \right)=\left(
                                         \begin{array}{c}
                                           \cos\frac{\theta}{2}\ket{00}-\sin\frac{\theta}{2}e^{-i\varphi}\ket{11} \\
                                           \cos\frac{\theta}{2}\ket{01}-i\sin\frac{\theta}{2}\ket{10} \\
                                           -i\sin\frac{\theta}{2}\ket{01}+\cos\frac{\theta}{2}\ket{10} \\
                                           \sin\frac{\theta}{2}e^{i\varphi}\ket{00}+\cos\frac{\theta}{2}\ket{11} \\
                                         \end{array}
                                       \right).
\end{eqnarray*}
Let us find the entanglement degree of the above states by using
negativity. For a pure two qubit state,
$\ket{\psi}=a\ket{00}+b\ket{11}$ or
$\ket{\phi}=a\ket{01}+b\ket{10}$, the negativity can be find to be
$N(\ket{\psi})=N(\ket{\phi})=2|ab|$. We can easily obtain the
negativity for the above entangled states as $N(\ket{e_{i}})
=|\sin\theta|$, where $i=1,2,3,4$. With the Yang-Baxter $\breve{R}$
acting on the standard bases, we can obtain a set of entangled
states, and these states possess the same entanglement degree which
depends on the parameter $\theta$. This character of the Yang-Baxter
$\breve{R}$ matrices has revealed in the Refs.. For the 2-qubit
quantum system, there is good entanglement measure
concurrence\cite{wootters1,wootters2}, $C(\rho_{12})=Max\{0,
\lambda_{1}-\lambda_{2}-\lambda_{3}-\lambda_{4}\}$. Here
$\{\lambda_{i}\}$ denotes the eigenvalues of the matrix
$\rho_{12}\sigma_{1}^{y}\sigma_{2}^{y}\rho_{12}^{*}\sigma_{1}^{y}\sigma_{2}^{y}$.
The notations $\rho_{12}$ and $\rho_{12}^{*}$ are biqubit density
matrix and its complex conjugate, correspondingly. The notations
$\sigma_{1,2}^{y}$ are pauli matrices. We can verify that
concurrence is equivalence to negativity for two-qubit ``X"
state(which density matrices are ``X" form).

\subsection{The $6 \times 6$ ``X" form $\breve{R}$ matrix}
When $j_{1}=1$ and $j_{2}=1/2$, with the relations in
Eqs.~(\ref{m-relation}), we can determine two matrices $M^{1,1/2}$
and $M^{1/2,1}$. In this section, the bases for the tensor product
space $V^{j_{1}}\otimes V^{j_{2}}$ are given by
$\{\ket{a\alpha}:a=1,0,-1;\alpha=1/2,-1/2\}$. In this case, the
Eqs.~(\ref{cond1}) gives the following relation,
\begin{equation}\label{6x6con}
    2\varphi_{0,1/2}=\varphi_{1,1/2}-\varphi_{1,-1/2}.
\end{equation}
If we set $\varphi_{1,1/2}=\varphi_{1}$ and
$\varphi_{1,-1/2}=\varphi_{2}$, then
$\varphi_{0,1/2}=(\varphi_{1}-\varphi_{2})/2$. Then a 6-dimensional
$M^{1,1/2}$ matrix is given as follows,
\begin{eqnarray}
  M^{1,1/2}&=&(e^{-i\varphi_{1}}\ket{1,1/2}\bra{-1,-1/2}+e^{-i\varphi_{2}}\ket{1,-1/2}\bra{-1,1/2}\nonumber\\
  &+&e^{-i(\varphi_{1}-\varphi_{2})}\ket{0,1/2}\bra{0,-1/2})+H.C
\end{eqnarray}
The $M^{1,1/2}$ matrix takes the following matrix form,
\begin{eqnarray}
 M^{1,1/2}=\left(
                                 \begin{array}{cccccc}
                                   0 & 0 & 0 & 0 & 0 & e^{-i\varphi_{1}} \\
                                   0 & 0 & 0 & 0 & e^{-i\varphi_{2}} & 0 \\
                                   0 & 0 & 0 & e^{-i(\varphi_{1}-\varphi_{2})} & 0 & 0 \\
                                   0 & 0 & e^{i(\varphi_{1}-\varphi_{2})} & 0 & 0 & 0 \\
                                   0 & e^{i\varphi_{2}} & 0 & 0 &0 & 0 \\
                                   e^{i\varphi_{1}} & 0 & 0 & 0 & 0 & 0 \\
                                 \end{array}
                               \right)
\end{eqnarray}
Then a 6-dimensional Yang-Baxter $\breve{R}^{1,1/2}(\theta)$ can be
construct as following,
\begin{equation}\label{6x6Rmatrix}
    \breve{R}^{1,1/2}(\theta)=\cos
    \frac{\theta}{2}I^{1,1/2}-i\sin\frac{\theta}{2}M^{1,1/2}
\end{equation}
When $\breve{R}^{1,1/2}(\theta)$ act on the standard basis(product
states),
\begin{eqnarray}
  \left(
     \begin{array}{l}
       \ket{e_{1}} \\
       \ket{e_{2}} \\
       \ket{e_{3}} \\
       \ket{e_{4}} \\
       \ket{e_{5}} \\
       \ket{e_{6}} \\
     \end{array}
   \right)
   = \breve{R}^{1,1/2}(\theta)\left(
     \begin{array}{l}
       \ket{1,1/2} \\
       \ket{1,-1/2} \\
       \ket{0,1/2} \\
       \ket{0,-1/2} \\
       \ket{-1,1/2} \\
       \ket{-1,-1/2} \\
     \end{array}
   \right)
\end{eqnarray}
Then we obtain six entangled states,
\begin{eqnarray}\label{6x6states}
  \begin{array}{l}
    \ket{e_{1}}=\cos\frac{\theta}{2}\ket{1,1/2}-i\sin\frac{\theta}{2}e^{-i\varphi_{1}}\ket{-1,-1/2} \\
    \ket{e_{2}}=\cos\frac{\theta}{2}\ket{1,-1/2}-i\sin\frac{\theta}{2}e^{-i\varphi_{2}}\ket{-1,1/2} \\
    \ket{e_{3}}=\cos\frac{\theta}{2}\ket{0,1/2}-i\sin\frac{\theta}{2}e^{-i(\varphi_{1}-\varphi_{2})}\ket{0,-1/2} \\
    \ket{e_{4}}=-i\sin\frac{\theta}{2}e^{i(\varphi_{1}-\varphi_{2})}\ket{0,1/2}+\cos\frac{\theta}{2}\ket{0,-1/2} \\
    \ket{e_{5}}=-i\sin\frac{\theta}{2}e^{i\varphi_{2}}\ket{1,-1/2}+\cos\frac{\theta}{2}\ket{-1,1/2} \\
    \ket{e_{6}}=-i\sin\frac{\theta}{2}e^{i\varphi_{1}}\ket{1,1/2}+\cos\frac{\theta}{2}\ket{-1,-1/2}
  \end{array}
\end{eqnarray}
Using the formula of negativity, we can obtain the entanglement
degree for the eigenstates of this Yang-Baxter system as
$N(\ket{e_{i}})=|\sin\theta|$. These eigenstates possess the same
degree of entanglement.

\subsection{The 8$\times$8 Yang-Baxter system}
When $j_{1}=3/2$ and $j_{2}=1/2$, we can obtain a $8\times 8$
$M^{3/2,1/2}$ matrix which satisfying the relations
Eqs.(\ref{m-relation}). For the following convenience, we introduce
the notation $\{\ket{i}; i=1,2\cdots 8\}$ to denote the standard
three-qubit basis.
\begin{eqnarray*}
  M^{3/2,1/2}&=&i(e^{-i\varphi_{1}}s_{1}^{+}s_{2}^{+}s_{3}^{+}+e^{-i\varphi_{2}}s_{1}^{+}s_{2}^{+}s_{3}^{-}+e^{-i\varphi_{3}}s_{1}^{+}s_{2}^{-}s_{3}^{+}+e^{-i\varphi_{4}}s_{1}^{+}s_{2}^{-}s_{3}^{-})\\
  &-&i(e^{i\varphi_{4}}s_{1}^{-}s_{2}^{+}s_{3}^{+}+e^{i\varphi_{3}}s_{1}^{-}s_{2}^{+}s_{3}^{-}+e^{i\varphi_{2}}s_{1}^{-}s_{2}^{-}s_{3}^{+}+e^{i\varphi_{1}}s_{1}^{-}s_{2}^{-}s_{3}^{-})
\end{eqnarray*}
If parameters $\varphi_{i}$$^{,}s$ satisfy the relation
$\varphi_{1}+\varphi_{4}=\varphi_{2}+\varphi_{3}$, then the
$M^{\frac{3}{2}\frac{1}{2}}$ satisfy the relations in
Eqs.(\ref{m-relation}).
 Then we can obtain a $8\times 8$ unitary
Yang-Baxter $\breve{R}-$matrix,
\begin{eqnarray*}
  \breve{R}^{3/2,1/2}(\theta) = \cos \frac{\theta}{2} I^{3/2,1/2}-i\sin
  \frac{\theta}{2}M^{3/2,1/2}
\end{eqnarray*}
We can verify that the Yang-Baxter $\breve{R}^{3/2,1/2}(\theta)$
matrix is unitary(\emph{i.e.}
$\breve{R}(\theta)^{\dag}\breve{R}(\theta)=\breve{R}(\theta)\breve{R}(\theta)^{\dag}=I$).
Let $H_{0}=s_{1}^{3}\otimes I_{2}\otimes I_{3}$.
 With this Yang-Baxter $\breve{R}-$matrix and this simple Hamiltonian, we can derive a
hamiltonian as $H
=\breve{R}(\theta)^{\dag}H_{0}\breve{R}(\theta)=\sum_{i=1}^{4}\textbf{B}_{i}\cdot
\textbf{S}_{i}$, where $\textbf{B}_{i}=(sin\theta cos\varphi_{i},
sin\theta sin\varphi_{i}, cos\theta)$ and
\begin{eqnarray*}
\begin{array}{ccc}
  S_{1}^{+}=\ket{1}\bra{8}, &  S_{1}^{-}=\ket{8}\bra{1}, &
  S_{1}^{3}=\frac{1}{2}(\ket{1}\bra{1}-\ket{8}\bra{8});
  \\
 S_{2}^{+}=\ket{2}\bra{7}, &  S_{2}^{-}=\ket{7}\bra{2}, &
 S_{2}^{3}=\frac{1}{2}(\ket{2}\bra{2}-\ket{7}\bra{7});
 \\
  S_{3}^{+}=\ket{3}\bra{6}, &  S_{3}^{-}=\ket{6}\bra{3}, &
  S_{3}^{3}=\frac{1}{2}(\ket{3}\bra{3}-\ket{6}\bra{6});
  \\
  S_{4}^{+}=\ket{4}\bra{5}, &  S_{4}^{-}=\ket{5}\bra{4}, &
  S_{4}^{3}=\frac{1}{2}(\ket{4}\bra{4}-\ket{5}\bra{5}).
\end{array}
\end{eqnarray*}

After some algebra, we can obtain the eigenvalues
$\{E_{i}^{\alpha}\}$ and eigenvectors $\{\ket{e_{i}^{\alpha}}\}$
$(\alpha=+,-; i=1,2,3,4)$ for Hamiltonian $H$ as following,
$$E_{i}^{+}=-E_{i}^{-}=1/2,$$
and corresponding eigenvectors,
\begin{eqnarray*}
\begin{array}{ccc}
  \ket{e_{1}^{+}}=cos\frac{\theta}{2}\ket{1}+sin\frac{\theta}{2}e^{i\varphi_{1}}\ket{8}, &
  \ket{e_{1}^{-}}=-sin\frac{\theta}{2}e^{-i\varphi_{1}}\ket{1}+cos\frac{\theta}{2}\ket{8};
  \\
 \ket{e_{2}^{+}}=cos\frac{\theta}{2}\ket{2}+sin\frac{\theta}{2}e^{i\varphi_{2}}\ket{7}, &
 \ket{e_{2}^{-}}=-sin\frac{\theta}{2}e^{-i\varphi_{2}}\ket{2}+cos\frac{\theta}{2}\ket{7};
 \\
  \ket{e_{3}^{+}}=cos\frac{\theta}{2}\ket{3}+sin\frac{\theta}{2}e^{i\varphi_{3}}\ket{6}, &
  \ket{e_{3}^{-}}=-sin\frac{\theta}{2}e^{-i\varphi_{3}}\ket{3}+cos\frac{\theta}{2}\ket{6};
  \\
  \ket{e_{4}^{+}}=cos\frac{\theta}{2}\ket{4}+sin\frac{\theta}{2}e^{i\varphi_{4}}\ket{5}, &
  \ket{e_{4}^{-}}=-sin\frac{\theta}{2}e^{-i\varphi_{4}}\ket{4}+cos\frac{\theta}{2}\ket{5}.
\end{array}
\end{eqnarray*}
In fact, the Hamiltonian $H$ can be recast as following,
\begin{eqnarray}\label{yangianH}
   H&=&\sum_{i=1}^{4}(\ket{e_{i}^{+}}\bra{e_{i}^{+}}-\ket{e_{i}^{-}}\bra{e_{i}^{-}})
\end{eqnarray}
Consider the state $|\psi\rangle$ in a three-qubit Hilbert space
$|\psi\rangle \in {\cal H}_A\otimes{\cal H}_B\otimes{\cal H}_C$. Its
coefficients with respect to a basis of product states (the
`computational basis') are $\psi_{i}=\langle i|\psi\rangle$, $i\in
\{0, 1 \cdots 8\}$. An important measure for the entanglement in
pure three-qubit states is the three-tangle (or residual tangle)
introduced in Ref.\cite{wootters3}. The three-tangle of
$|\psi\rangle$ is a so-called polynomial invariant and  can be
written in terms of the coefficients $\psi_{i}$ as
\begin{eqnarray}
  \label{eq:3tangle}
  \tau_3(\psi)\  &=&\  4|d_1-2d_2+4d_3|\\
  d_1\  &=&\  \psi_{1}^2\psi_{8}^2 + \psi_{2}^2\psi_{7}^2 + \psi_{3}^2\psi_{6}^2
  + \psi_{5}^2\psi_{4}^2\nonumber\\
  d_2\  &=&\  \psi_{1}\psi_{8}\psi_{4}\psi_{5} +
  \psi_{1}\psi_{8}\psi_{6}\psi_{3} + \psi_{1}\psi_{8}\psi_{7}\psi_{2} \nonumber\\
  && {}+\psi_{4}\psi_{5}\psi_{6}\psi_{3} + \psi_{4}\psi_{5}\psi_{7}\psi_{2} +
  \psi_{6}\psi_{3}\psi_{7}\psi_{2}\nonumber\\
  d_3\  &=&\  \psi_{1}\psi_{7}\psi_{6}\psi_{4} + \psi_{8}\psi_{2}\psi_{3}\psi_{5}\ \ .\nonumber
\end{eqnarray}
Then we can obtain three-tangle for the Eigenstates are as
following,
\begin{eqnarray*}
  \tau_{3}(\ket{e_{i}^{\alpha}}) &=& sin^{2}\theta.
\end{eqnarray*}
By using the definition of concurrence we can obtain the
\begin{eqnarray*}
  C_{AB}(\ket{e_{i}^{\alpha}})=C_{AC}(\ket{e_{i}^{\alpha}})=C_{BC}(\ket{e_{i}^{\alpha}})=0
\end{eqnarray*}
where $i=1,2,3,4$ and $\alpha=+,-$. When the parameter
$\theta=\pi/2$, $\tau_{3}(\ket{e_{i}^{\alpha}})=1$ and
$C_{XY}(\ket{e_{i}^{\alpha}})=0$($XY=AB,BC,AC$). Then we can say
these eigenstates are GHZ type states.

\section{\label{sec4}Yangian symmetry and shift operators}
In the Sec.\ref{sec3}, we construct a Hamiltonian(\emph{i.e.}
Eq.(\ref{yangianH})) with the Yang-Baxter $\breve{R}^{3/2,1/2}$
matrix. As is known to all, the Yangian is a very important tool to
study symmetry and shift operators. Motivated this, we will
investigate the symmetry to this Yang-Baxter Hamiltonian and Yangian
generators as shift operators in detail.

In fact, with the eigenvectors $\{\ket{e_{i}^{\alpha}}\}$ we can
construct a special Yangian Y(\emph{sl}(2)) realization $\{I_{\pm},
I_{3}\}$ and $\{F_{\pm}, F_{3}\}$ as following,
\begin{eqnarray*}
  I_{+}&=&\ket{e_{1}^{+}}\bra{e_{2}^{+}}+\ket{e_{3}^{+}}\bra{e_{4}^{+}}+\ket{e_{1}^{-}}\bra{e_{2}^{-}}+\ket{e_{3}^{-}}\bra{e_{4}^{-}} \\
  I_{-}&=&\ket{e_{2}^{+}}\bra{e_{1}^{+}}+\ket{e_{4}^{+}}\bra{e_{3}^{+}}+\ket{e_{2}^{-}}\bra{e_{1}^{-}}+\ket{e_{4}^{-}}\bra{e_{3}^{-}} \\
  I_{3}&=&
  \frac{1}{2}[(\ket{e_{1}^{+}}\bra{e_{1}^{+}}+\ket{e_{3}^{+}}\bra{e_{3}^{+}}+\ket{e_{1}^{-}}\bra{e_{1}^{-}}+\ket{e_{3}^{-}}\bra{e_{3}^{-}})\\
  &-&(\ket{e_{2}^{+}}\bra{e_{2}^{+}}+\ket{e_{4}^{+}}\bra{e_{4}^{+}})+\ket{e_{2}^{-}}\bra{e_{2}^{-}}+\ket{e_{4}^{-}}\bra{e_{4}^{-}})],
\end{eqnarray*}
and
\begin{eqnarray*}
  F_{+}&=&2\alpha(\ket{e_{1}^{+}}\bra{e_{4}^{+}}+\beta\ket{e_{3}^{+}}\bra{e_{2}^{+}})+2\gamma(\ket{e_{1}^{-}}\bra{e_{4}^{-}}+\delta\ket{e_{3}^{-}}\bra{e_{2}^{-}})\\
  F_{-}&=&2\alpha(\beta\ket{e_{4}^{+}}\bra{e_{1}^{+}}+\ket{e_{2}^{+}}\bra{e_{3}^{+}})+2\gamma(\delta\ket{e_{4}^{-}}\bra{e_{1}^{-}}+\ket{e_{2}^{-}}\bra{e_{3}^{-}})\\
  F_{3}&=&\alpha(\ket{e_{1}^{+}}\bra{e_{3}^{+}}-\ket{e_{2}^{+}}\bra{e_{4}^{+}}+\beta\ket{e_{3}^{+}}\bra{e_{1}^{+}}-\beta\ket{e_{4}^{+}}\bra{e_{2}^{+}})\\
 &&
 +\gamma(\ket{e_{1}^{-}}\bra{e_{3}^{-}}-\ket{e_{2}^{-}}\bra{e_{4}^{-}}+\delta\ket{e_{3}^{-}}\bra{e_{1}^{-}}-\delta\ket{e_{4}^{-}}\bra{e_{2}^{-}}).
\end{eqnarray*}

It is not difficulty to verify that $\{I_{\pm}, I_{3}\}$ and
$\{F_{\pm}, F_{3}\}$ satisfy the following Yanigian
$Y(\emph{sl}(2))$ relations,
\begin{eqnarray*}
   &&[I_{3},I_{\pm}]=\pm I_{\pm}, ~~[I_{+},I_{-}]=2 I_{3}\\
   &&[I_{3},F_{\pm}]=[F_{3},I_{\pm}]=\pm F_{\pm},~~~[I_{\pm},F_{\mp}]=\pm 2 F_{3}\\
   &&[I_{3},F_{3}]=[I_{\pm}, F_{\pm}]=0,
\end{eqnarray*}
and
\begin{eqnarray*}
  &&[F_{3},[F_{+},F_{-}]] =
  0,~~~[F_{\pm},[F_{3},F_{\pm}]] =
  0\\
  && [F_{\pm},
  [F_{\pm},F_{\mp}]]\pm 2[F_{3},[F_{3},F_{\pm}]] =0.
\end{eqnarray*}
We can verify that the Hamiltonian and Yangian operators satisfy the
following relation,
\begin{eqnarray*}
  [H,Y_{\alpha}] &=& 0,
\end{eqnarray*}
\begin{figure}[htbp]
\begin{center}
\includegraphics[width=6cm]{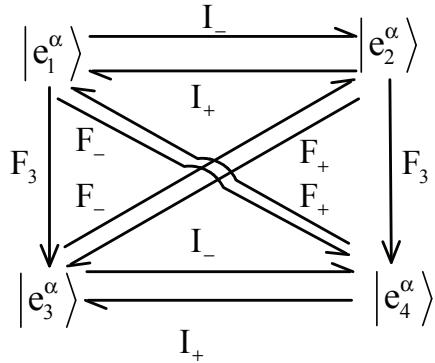}
\caption{The states transfer graph for the Yang-Baxter
Hamiltonian$(\alpha=\pm)$.}\label{fig}
\end{center}
\end{figure}
where $Y=I,F$ and $\alpha=\pm,3$. That is to say this Hamiltonian
possess a Yangian Y(\emph{sl}(2)) symmetry. This maybe the simplest
Hamiltonian with Yangian $Y(\emph{sl}(2))$ symmetry. In quantum
physics, the Yangian generators can be used to construct shift
operators. Then we will construct shift operators for this
Yang-Baxter Hamiltonian. When the Yangian operators $\{I_{\pm},
I_{3}\}$ and $\{F_{\pm}, F_{3}\}$ act on the eigenstates of this
Yang-Baxter Hamiltonian, we can obtain a state transfer graph in
Fig.(\ref{fig}).

\section{Summary}

In this paper, we construct a set of
$(2j_{1}+1)\times(2j_{2}+1)-$dimensional ``X" form Yang-Baxter
$\breve{R}^{j_{1}j_{2}}(\theta)$. We investigated this set unitary
Yang-Baxter $\breve{R}^{j_{1}j_{2}}(\theta)$ as quantum gate in
quantum computation processing. When these ``X" form Yang-Baxter
$\breve{R}^{j_{1}j_{2}}(\theta)$ matrices act on standard bases, we
can obtain a set of entangled states, which possess the same degree
of quantum entanglement. We also construct a Yang-Baxter Hamiltonian
with Yangian Y(\emph{sl(2)}) symmetry. And Yangian generators can be
viewed as shift operators.

\section*{Acknowledgments}
This work was supported by NSF of China (Grants No. 10875026) and
the Fundamental Research Funds for the Central Universities(Grants
No. 09SSXT026)


\begin{thebibliography}{}
\bibitem{ben1} C. H. Bennett and D. P. DiVincenzo.:Quantum information and computation. Nature \textbf{404}
247(2000).
\bibitem{ben2} C. H. Bennett and G. Brassard, C. Cr\'{e}peau, R. Jozsa, A Peres, and W. K. Wootters.:Teleporting
an unknown quantum state via dual classical and
Einstein-Podolsky-Rosen channels. Phys. Rev. Lett. \textbf{70},
1895(1993).
\bibitem{ben3} C H. Bennett and S. J. Wiesner.:Communication via one- and two-particle operators on
Einstein-Podolsky-Rosen states. Phys. Rev. Lett. \textbf{69},
2881(1992).
\bibitem{murao} M. Murao, D. Jonathan, M. B. Plenio, and V. Vedral.:Quantum telecloning and multiparticle
entanglement, Phys. Rev. \textbf{A 59}, 156(1999).
\bibitem{nielsen}M. Nielsen and I. Chuang.: Quantum Computation and Quantum Information, Cambridge University
Press(2000)
\bibitem{nayak}Sankar Das Sarma, Michael Freedman, and Chetan Nayak.:Topologically Protected Qubits from a Possible Non-Abelian Fractional Quantum Hall State, Phys. Rev. Lett. \textbf{94},166802(2005).

\bibitem{kauffman1}L. H. Kauffman.: Knots and Physics, World Scientific Publishers(2002).
\bibitem{kauffman2}L. H. Kauffman and S. J. Lomonaco Jr.:Braiding operators are universal quantum gates. New J. Phys.\textbf{4},73.1每73.18.(2002).
\bibitem{zhang1} Yong Zhang,Louis H. Kauffman, and Mo-Lin Ge.:Yang每Baxterizations, Universal Quantum Gates and Hamiltonians, Quantum Information Processing, Vol. \textbf{4}, No. 3, August
(2005).
\bibitem{yang} C. N. Yang.: Some Exact results for the many-body problem in one dimension with
repulsive delta-function interaction. Phys. Rev. Lett. \textbf{19},
1312(1967); C. N. Yang.: S matrix for the one-dimensional N-body
problem with repulsive or attractive -function interaction. Phys.
Rev. \textbf{168} 1920(1968).
\bibitem{baxter} R. J. Baxter.:Exactly Solved Models in Statistical Mechanics Academic
Press, London, (1982); R. J. Baxter.:Partition funtion of the
eighy-vertex lattice model. Ann. Phys. \textbf{70}, 193(1972).
\bibitem{zhang2} Y. Zhang, L. H. Kauffman, and M. L. Ge.: Universal quantum gate, YangBaxterization
and Hamiltonian. Int. J. Quant. Inf.\textbf{3} 669(2005).
\bibitem{chen1} J. L. Chen, K. Xue, and M. L. Ge.: Braiding transformation, entanglement swapping, and
Berry phase in entanglement space. Phys. Rev. A. \textbf{76},
042324(2007).
\bibitem{chen2} J. L. Chen, K. Xue, and M. L. Ge.: Berry phase and quantum criticality in Yang Baxter
systems. Ann. Phys. \textbf{323} 2614(2008).
\bibitem{chen3} J. L. Chen, K. Xue, and M. L. Ge.: All pure two-qudit entangled states can be generated via
a universal Yang每Baxter matrix assisted by local unitary
transformations. Chinese Phys. Lett. \textbf{26}, 080306 (2009).
\bibitem{hu1}Shuang-Wei Hu,Kang Xue, and Mo-Lin Ge.: Optical simulation of the Yang-Baxter equationPhys. Rev. A
\textbf{78}, 022319(2008).
\bibitem{hu2} Ming-Guang Hu,Kang Xue, and Mo-Lin Ge.: Exact Solution of a Yang-Baxter Spin-1/2 Chain Model and Quantum Entanglement. Phys. Rev. A \textbf{78}, 052324 (2008)
\bibitem{wang1} Gangcheng Wang, Kang Xue, Chunfeng Wu, He Liang and C H
Oh.: Entanglement and the Berry phase in a new Yang-Baxter system.
J. Phys. A: Math. Theor. \textbf{42}, 125207(2009).
\bibitem{drin} V. G. Drinfeld.: Hopf algebras and the quantum Yang-Baxter equation. Soviet Math.  Dokl \textbf{32},pp. 254-258(1985).
\bibitem{xue1}C.M.Bai, M.L.Ge and K.Xue.:Yangian and its applications, Inspired by s.s. chen: A Memorial vol.II in Honor of A Great Mathematician, Edited by P.A. Griffiths, World Scientific, Singapore, 45-93(2006)
\bibitem{xue2}L.J.Tian, H.B.Zhang, S.Jin, K.Xue.:Y(sl(2)) algebra application in extended hydrogen atom and monopole models,Commun.Theor.,Phys.\textbf{41}(2004)641
\bibitem{xue3}M.L.Ge,L.C.Kwek, C.H.Oh, K.Xue.:Yangians and transition operators, Czech. J. phys. \textbf{50},1229(2000)
\bibitem{xue4} M.L. Ge, K. Xue and Y-S. Wu.: Explicit
Trigonometric Yang-Baxterization. Int. J. Mod. Phys. \textbf{A6},
3735(1991);
\bibitem{xue5}Y. Cheng, M.L. Ge and K. Xue, Yang Baxterization of Braid Group Repre-
sentations, Commun. Math. Phys. \textbf{136},195(1991).
\bibitem{xue6}M.L. Ge, Y.S. Wu and K. Xue, Explicit Trigonometric Yang每Baxterization,
Int. J. Mod. Phys A,\textbf{6},3735(1991)
\bibitem{zycz}K. Zyczkowski, \emph{et al.}: Volume of the set of separable states. Phys. Rev. A, \textbf{58}, 883, (1998).
\bibitem{wangxg}Xiaoguang Wang \emph{et al.}.: Negativity, entanglement witnesses and quantum phase transition in spin-1 Heisenberg chains.  J. Phys. A: Math. Theor. \textbf{40}
10759-10767(2007)
\bibitem{wootters1}Hill S,Wootters W.K. Entanglement of a pair of quantumbits.
Physical Review Letters, \textbf{78}, 5022每5025(1997),
\bibitem{wootters2}Wootters W.K. Entanglement of formation of an arbitrary state of two qubits.
Physical Review Letters,\textbf{ 80}, 2245每 2248(1998)
\bibitem{wootters3}V. Coffman, J. Kundu, and W. K. Wootters.:Phys. Rev. A \textbf{61}, 052306 (2000).


\end{thebibliography}


\end{document}